\documentclass[aps,pre,twocolumn,unsortedaddress]{revtex4-2}
\usepackage[utf8]{inputenc}
\usepackage{graphicx}
\usepackage{color}

\begin{document}
\title{Flow induced rigidity percolation in shear thickening suspensions}
\author{Abhay Goyal}
\author{Nicos S. Martys}
\affiliation{Infrastructure Materials Group, Engineering Laboratory, National Institute of Standards and Technology}
\author{Emanuela Del Gado}
\affiliation{Dept of Physics, Institute of Soft Matter Synthesis and Metrology, Georgetown University}

\begin{abstract}
Discontinuous shear thickening (DST) is associated with a sharp rise of a suspension's viscosity with increasing applied shear rate. A key signature of DST, highlighted in recent studies, is the very large fluctuations of the measured stress as the suspension thickens. A clear link between microstructural development and the dramatic increase of the stress fluctuations has not been established yet. To identify the microstructural underpinnings of this behavior, we perform simulations of sheared dense suspensions. By analyzing particle contact networks, we identify a subset of constrained particles that contribute directly to the rapid rise in viscosity and the large stress fluctuations. Indeed, both phenomena can be explained by the growth and percolation of constrained particle networks---in direct analogy to rigidity percolation. A finite size scaling analysis confirms this is a percolation phenomenon and allows us to estimate the critical exponents. Our findings reveal the specific microstructural transition that underlies DST. 
\end{abstract}

\maketitle



Suspensions play an important role in a wide variety of environmental and technological processes. Examples include colloidal systems, pharmaceuticals, slurries, and concrete. Their flows raise a number of fundamental physics questions, exhibiting a multitude of phenomena that include shear thinning and thickening, thixotropy, giant stress fluctuations and jamming \cite{Morris-Guazzelli,Mewis-Wagner}. While several problems remain outstanding, intense interest has focused on the physical mechanisms at the origin of the discontinuous shear thickening (DST), 
which is quite ubiquitous and dramatic: the suspension experiences a rapid rise in stress or viscosity as the imposed shear rate increases. 
The flow of DST suspensions rapidly becomes strongly dilatant \cite{Hoffman1972discontinuous,leighton_acrivos_1987,Royer2016} and erratic, with giant stress fluctuations, rapid increase of stress, and structural inhomogeneities in response to the increasing shear rate \cite{Lootens2003giant, Hermes2016unsteady,SaintMichel2018uncovering, Rathee2017localized}.
 \begin{figure*}
    \centering
    \includegraphics[width=.99\textwidth]{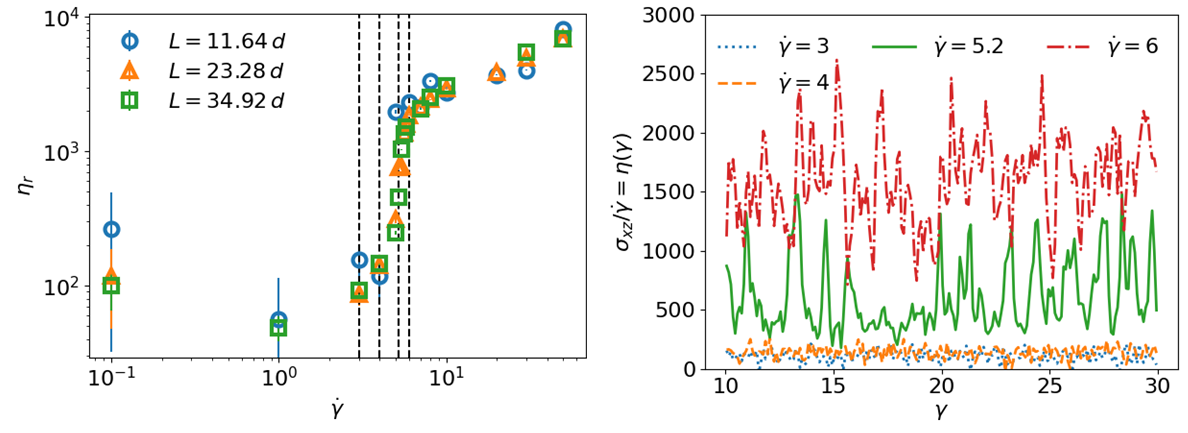}
    \caption{\textbf{(a)} The average relative viscosity ($\eta_r$) as a function of shear rate ($\dot{\gamma}$) for a suspension at $\phi=56\ \%$. The error bars represent one standard deviation from the mean that was calculated, after a dynamical equilibrium was reached, from a running average over 25 strain units. At low rates, there is a slight shear thinning effect. As the rate increases, we observe a sharp jump in the viscosity around $\dot{\gamma}=5\ {\tau_{0}}^{-1}$ corresponding to DST. Above this DST regime, there appears to be some continuous shear thickening (CST). However, tests with an increased $\eta_0$ show that the CST at high rates is replaced with a plateau in the viscosity if the Reynolds number is low. Regardless, the change in Reynolds number did not noticeably affect the DST transition region, in either critical shear rate or viscosity magnitude for the systems studied. The vertical dashed lines indicate the shear rates used in \textbf{(b)}, where we show the instantaneous viscosity (stress/shear rate) as a function of applied strain $\gamma$. The jump in $\eta_r$ is accompanied by large fluctuations of the stress near the critical shear rate. }
    \label{fig:viscosity}
\end{figure*}
 
 Recent studies on DST have gained new insights into the role of frictional solid-on-solid contacts between the particles in a suspension, and how their sharp increase in number largely controls the emergence of DST \cite{Seto2013,Mari2014,Wyart2014,Lin2015hydrodynamic}. Indeed, over the last few years, experiments and numerical simulations have extensively analyzed and demonstrated, how, at the onset of DST, the bulk rheological behavior of suspensions becomes dramatically sensitive to the surface interactions, roughness, and hence frictional contact between particles, in spite of the presence of solvent lubrication forces \cite{Lootens2004,Royer2016,Morris2018lubricated,Jamali2019,Wang2020,Ong2020,More2020,Nabizadeh2022structure}. This fundamental understanding of such phenomena has opened new paths to design the flow of dense suspensions through nanoscale physics and surface chemistry. However, in spite of the sensitivity to particle surface contacts which are strongly material and chemistry dependent, the overall DST phenomenology is consistent across the whole spectrum of suspensions involved, suggesting the presence of common microstructural features which remain elusive. 
 
 The theoretical mean-field approach of Wyart and Cates \cite{Wyart2014}, in which the microstructure is characterized by the suspension volume fraction and the overall fraction of frictional contacts produced under shear, demonstrated that the sudden rise in the stress (or decrease in shear rate for the case of stress driven shear) has general features that do not depend on the specifics of the material and surface chemistry. Further, the large fluctuations of the shear stresses 
 and the scaling properties of the shear response, which are reminiscent of critical phenomena, seem ubiquitous in DST over a wide range of suspensions \cite{Hermes2016unsteady,Rathee2017localized,SaintMichel2018uncovering,Sedes2020fluctuations,Ramaswamy2021}. 
 
 These findings suggest that, as frictional contacts between the particles become prevalent with increasing rate, larger scale microstructures, involving many particles and built under shear, may emerge independently from the detailed material chemistry and control the DST phenomenon \cite{Seto2019}. Previous studies have focused on order-disorder transitions in the microstructure of particle suspensions approaching shear thickening \cite{Bender1996,Kulkarni2009,Goyal2022}, and there is now growing consensus, from experiments and simulations, that large clusters or chains of particles spanning the system may cause the abrupt increase of stress or possibly jamming \cite{Seto2013,Wyart2014,Thomas2018microscopic,Edens2021,Gameiro2020interaction,Lin2016tunable}. Clearly, the fact that there is microstructural reorganization that eventually spans the system is suggestive of a percolation transition \cite{Stauffer2018,Sedes2022kcore}. The occurrence of a percolation transition could provide a connection to critical phenomena \cite{Stauffer2018}, and could provide a conceptual framework for understanding aspects of both the microstructural development and its link to DST. However, testing these ideas is extremely challenging, since microstructures, and in particular stress-bearing ones, are hardly accessible in experiments on dense suspensions subjected to high shear stresses or rates. Computer simulations of particle based models may instead be specifically designed to investigate those microstructures, and are therefore the most promising tool to shed light onto these questions. Nevertheless, due to their complexity, most simulations studies have been so far limited to relatively small system sizes, whereas much larger system sizes would be needed to detect critical-like behaviors, and identify their origin.     

Here we use large scale 3D simulations of model dense suspensions to show that a percolation transition is indeed at the origin of the stress fluctuations characteristic of DST, once one properly defines the basic unit of microstructure that forms the percolating structure. The percolation of this microstructure can be directly linked to the DST, and to the accompanying large stress fluctuations, and points to the role of rigidity percolation in this phenomenon. Further, the critical behavior is studied by applying a finite size scaling analysis, which allows us to estimate the related critical exponents. 

We have utilized computer simulations to study a model suspension of 
spheres that interact via hydrodynamic lubrication, contact repulsion, and frictional forces, following recent work on simulations of shear thickening suspensions \cite{Mari2015,Singh2018,Goyal2022}. All spheres have the same size, while all interaction parameters, provided in the supplementary information (SI), have been adjusted to match the model in \cite{Singh2018} where lubrication forces are regularized at short distances between the particles surfaces and Coulomb friction act tangentially on surface contacts. We follow the same numerical approach described in \cite{Goyal2022} to integrate the equations of motion for all particles, which allows us to perform simulations of large systems with LAMMPS \cite{Plimpton1995} with overdamped particle motions. The system is sheared using Lees-Edwards boundary conditions, and a background velocity field with constant shear rate is imposed that particle motion can relax to. The system is also subject to thermal fluctuations. All quantities are reported in reduced units as a combination of three basic units: energy scale $\varepsilon=k_BT$, particle mass $m$, and particle diameter $d$. From these parameters, the unit of time is $\tau_{0}=\sqrt{md^2/\varepsilon}$. Here we present data on volume fraction $\phi=56 \, \% $, and we reproduce, as in \cite{Singh2018}, the features of the DST. 
The data reported in the following refer to simulation boxes with edge length, $L$= 11.64 $d$, 23.28 $d$ and 34.92 $d$, containing 1688, 13500 and 45563 spheres respectively. The medium and larger sizes correspond to much larger system sizes than previous studies, which is essential to addressing the critical-like behavior of stress fluctuations and to deal with finite-size effects. We use the convention that $x$ corresponds to the flow direction, $y$ corresponds to the vorticity direction, and $z$ to the shear gradient direction. The model suspensions were sheared up to 40 strain units to reach a steady flow state.     
\begin{figure}
    \centering
    \includegraphics[width=.49\textwidth]{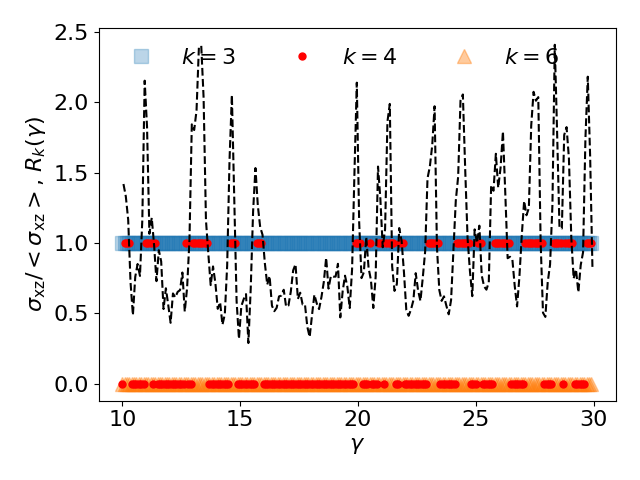}
    \caption{We characterize the percolation of the $k$-neighbor particle network with a parameter $R_k(\gamma)$, which is equal to 1 if there exists at a percolating cluster of $k$-neighbor particles and 0 if there does not. Hypothesizing that there is some value of $k$, potentially given by the Maxwell criterion of $N_{\rm dim}+1$, for which particles are sufficiently constrained to support stresses, we plot $R_k(\gamma)$ for different $k$ values alongside the stress $\sigma_{\rm xz}/<\sigma_{\rm xz}>$ (dashed black line). While considering $k\leq3$, there is always a percolating cluster near the critical rate, and there seems to be no connection between the percolation and stress fluctuation. Conversely, with larger value of $k=6$, no percolation is observed. However, when we select $k=4$, we identify a strong correlation between the shear stress and the percolation of $k$neighbor particles.}
    \label{fig:kcorePercolation}
\end{figure}
We use the shear component $\sigma_{xz}$ of the stress tensor, obtained from interparticle forces, relative positions, and particle velocities \cite{Evans1990}, to extract the relative viscosity $\eta_r$ which is plotted in Figure \ref{fig:viscosity}a as a function of the shear rate $\dot{\gamma}$ for the three system sizes studied. The data show a steep increase of viscosity at a threshold shear rate $\dot{\gamma} \simeq 5\  {\tau_{0}}^{-1}$, and increasing the system size does not significantly alter the viscosity increase, indicating that the DST phenomenon identified here is not an artifact of limited sample sizes used in simulations. The time series from which the viscosity is extracted in steady state are plotted in Figure \ref{fig:viscosity}b, which shows (for the intermediate system size of 13500 particles) $\sigma_{xz}/\dot{\gamma}$ as a function of the strain $\gamma$ for several shear rates. The data clearly demonstrate the presence of large fluctuations of the shear stress in systems that have gone through DST. 

As described earlier, the number of frictional contacts per particle is strongly correlated to the suspension stress close to shear thickening. Recent work \cite{Sedes2022kcore} has shown that the shear stress increases with the mean frictional contact number, peaking at 3 and 4 for large shear rates. However, in terms of the mean frictional contact number, a distinction could not be made between continuous shear thickening (CST) and DST, which clearly have different rheological signatures.
These findings point, in our view, to the fact that the local microstructural environment and its larger scale connectivity determines the nature of the stress transmission through the particle suspension. As a consequence, to search for the microscopic origin of the stress fluctuations, we consider the hierarchy of structures built up by particles that share frictional contacts with a minimum $k$ neighbors.
In 
graph theory this defines a contact graph (also called a network) composed of nodes with degree $k$ at minimum \cite{Morone2019jamming,Papadopoulos2018network}.
To be more precise, we define a neighbor, for each particle, as a particle close enough in proximity that it can interact with said particle by frictional forces. In our frictional model, this occurs when the sphere surfaces are in contact. The frictional force is proportional to the normal force between the neighboring spheres and acts to resist transverse motion relative to surface normal between neighboring spheres (see SI). We will refer to particles with at least $k$ neighbors as $k$-neighbor particles.

\begin{figure}
    \centering
    \includegraphics[width=.49\textwidth]{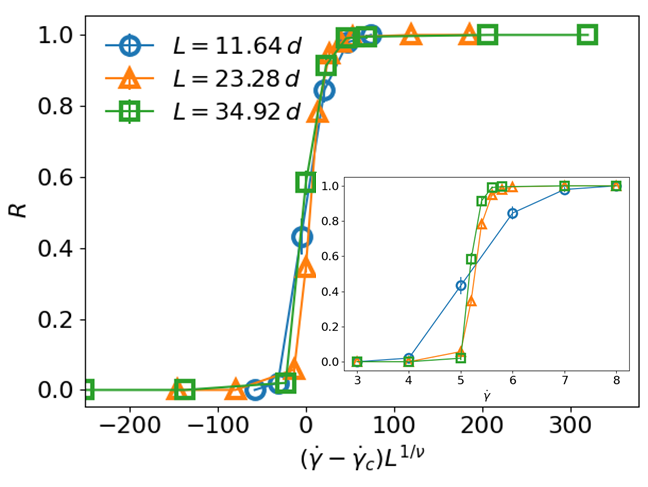}
    \caption{For each system size and shear rate, we can compute the probability of $k$-neighbor particles forming a percolating cluster. This is plotted for against the shear rate for three system sizes.   There is a clear percolation threshold at about $\dot{\gamma}=5.2 \ {\tau_{0}}^{-1}$, but due to finite size effects the curves are a bit distorted---especially at the smallest system size (inset). However, following percolation theory, these can be rescaled by the system size raised to a certain power (defined as $\nu$). We perform this rescaling and get a nice data collapse with $\nu\approx0.6$. Interestingly, this is smaller than the value for standard percolation in 3D ($\nu=0.88$). Note, the standard uncertainty in R is smaller than the shown symbols. Finally, we add that despite the imposed anisotropy due to the applied shear boundary condition, we did not find any significant difference in the percolation probabilities in the vorticity, flow, and shear gradient directions.}
    \label{fig:percolationProb}
\end{figure}
Testing $k$ values from 2 to 6, we 
determine whether the percolation of $k$-neighbor particles can be directly connected to the rapid rise in viscosity and the large stress fluctuations corresponding to the DST in our simulations. For each $k$, we identify the $k$-neighbor particles, sort them into clusters, 
obtain the cluster size distribution, and identify configurations in which at least one percolating cluster is present in all directions. The plot in Figure \ref{fig:kcorePercolation} superimposes the time series for the presence of a percolating cluster of $k$-neighbor particles ($R_{k}=1$) or not ($R_{k}=0$), for different $k$, to the time series of the shear stress close to DST. The data show that clusters of 3-neighbor particles always percolate, independently from the stress fluctuations, which also happens for $k$=2 (data not shown). In contrast, the percolation of a $4$-neighbor cluster exactly corresponds to the spikes in the shear stress of the suspension. Note, the $4$-neighbor  particles locally satisfies the Maxwell criterion for rigidity in presence of tangential frictional forces \cite{Liu2019frictional}. We also find that at low enough $\dot{\gamma}$, where DST does not occur and stress fluctuations are much smaller, the percolation of $4$-neighbor particles were not observed. At higher $k$ values, percolation is significantly reduced \cite{Sedes2022kcore}. Indeed, the data in Figure \ref{fig:kcorePercolation} show that percolation of $6$-neighbor particles, which, incidentally, correspond to locally rigid structures for frictionless spheres, is not evident.
These findings strongly suggest that the percolation of locally rigid structures, reminiscent of shear jamming in granular fluids \cite{Dapeng2011jamming,Vinutha2016}, may be at the origin of the large stress fluctuations typical of DST, playing a significant role in this phenomenon. Indeed, the percolation of the rigid 4-neighbor particles can support the transmission of stress over an extended period of time, in contrast to $k$-neighbor particles with $k$ lower values, which can more easily be disrupted. The percolation of the rigid 4-neighbor particles may therefore be central to the self-organization of the microstructure of the suspensions under flow, when DST occurs.

It is intriguing, at this point, to ask whether the growth of $4$-neighbor particles can be studied in terms of percolation theory. That is, can we define a critical shear rate and critical exponents that describe a diverging length scale and mean cluster size? The determination of such quantities can be challenging even for static systems. Here we use a finite size scaling ansatz typical of critical phenomena and percolation \cite{Stauffer2018} to first determine the critical shear rate. We examine the probability of percolation, $R$, defined as the average occurrence of percolating clusters of $4$-neighbor particles over each time series at different shear rates and for the 3 different sizes of the simulation box. The data for $R$ as a function of shear rate (Figure \ref{fig:percolationProb} inset) show that the larger the system size the steeper the transition from 0 to 1 in probability, akin to the behavior of the percolation probability close to a percolation transition, and that the curves for each system size intersect at around a shear rate of about 5. Hence we hypothesize that the percolation threshold of the $4$-neighbor particles indeed corresponds to a specific shear rate. By defining the approximate intersection of the different curves as the critical shear rate corresponding to the percolation threshold, we can collapse all data in terms of a scaling variable $(\dot {\gamma} - \dot {\gamma}_c )L^{1/\nu}$. ${\gamma}_c$ points to a characteristic time scale over which a stable percolating network may be built or destroyed at a given shear rate (Figure \ref{fig:percolationProb}). In general, we would expect that, quantitatively, this characteristic timescale depend on the microscopic physics of the system, and hence on the specific experimental system considered or on the microscopic parameters of the simulations (see SI). 

\begin{figure*}
    \centering
    \includegraphics[width=\textwidth]{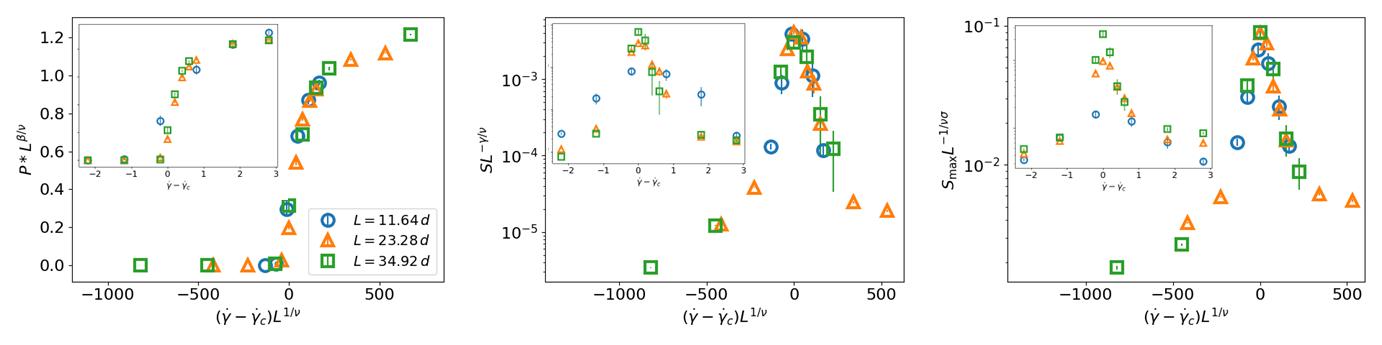}
    \caption{A number of quantities can be computed in a percolating system that show interesting behavior near the critical point, including the probability of a random $k$-neighbor particle being in the percolating cluster ($P$), the mean non-percolating cluster size ($S$), and the maximum non-percolating cluster size ($S_{\rm max}$). Each of these quantities are affected by the finite system size (inset), but they can be rescaled to produce a data collapse as shown in the main figures. We obtain an initial estimate of the three exponents: $\beta\approx 0.18$, $\sigma\approx 0.75$, and $\gamma\approx 1.3$.}
    \label{fig:sizeScaling}
\end{figure*}

The data collapse supports the validity of the finite size scaling ansatz and provides a first estimate for the critical exponent $\nu$, $ \nu \approx 0.6$, which describes the divergence of the correlation length close to the percolation threshold. We can then compute, from the same cluster analysis of $4$-neighbor particles and having obtained the whole cluster size distribution, the probability of a $k$-neighbor random particle being in the percolating cluster, the mean cluster size (i.e. the second moment of the cluster size distribution), and the maximum cluster size \cite{Stauffer2018}. All these quantities follow the data collapse which stems from the finite size scaling ansatz (Figure \ref{fig:sizeScaling}), and the collapsed data allow us to independently determine three different critical exponents: $\beta\approx 0.18$, $\sigma\approx 0.75$, and $\gamma\approx 1.3$. The exact determination of the critical exponents will require further studies, however, the data already show a significant discrepancy from the mean field values and from the random connectivity percolation transition in 3$D$, suggesting that the percolation of the $4$-neighbor particles during DST may correspond to a distinct universality class. Rigidity percolation studies, while quite limited (especially in 3$D$), have indeed suggested a distinct universality class, and frictional rigidity percolation studies have found similar discrepancies \cite{jakob1995generic2d,Liu2019frictional}.    








\begin{figure*}
    \centering
    \includegraphics[width=.8\textwidth]{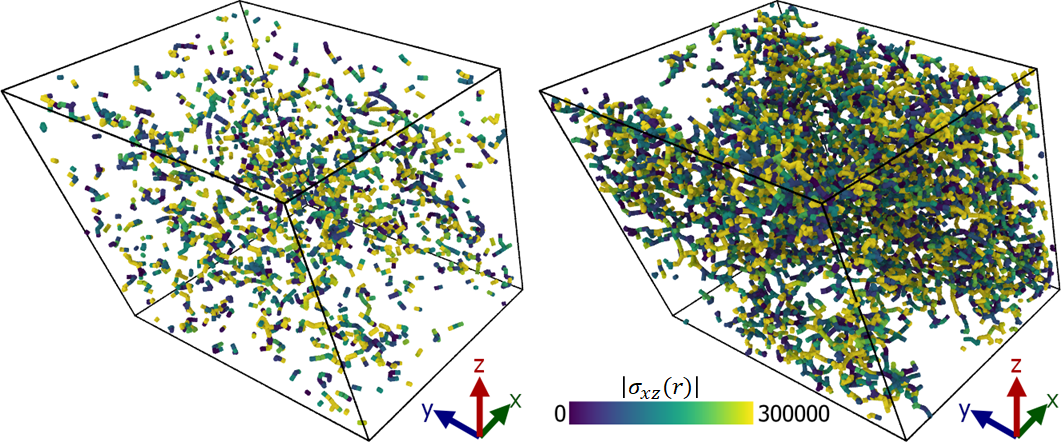}
    \caption{Snapshots showing the connections between $4$-neighbor particles at $\dot{\gamma}=5\,{\tau_{0}}^{-1}$, but at different strains, $\gamma$. The connections are colored by the magnitude of the local shear stress, $| \sigma_{xz} |$. While both snapshots are at the same shear rate, the different $\gamma$ values correspond to very different $ \sigma_{xz}$. \textbf{Left:} At $\sigma_{xz} \approx 500 \, k_BT/d^3$, we show all $4$-neighbor particles. These are distributed homogeneously but do not percolate. \textbf{Right:} At a later point, the stress increases to $\sigma_{xz} \approx 4000 \, k_BT/d^3$, which is accompanied by a large increase in the number of $4$-neighbor particles and their percolation. Here we show only the percolating cluster, corresponding to roughly half of the $4$-neighbor particles present, which forms a densely connected network that spans the system in all directions. }
    \label{fig:visualization}
\end{figure*}

In conclusion, we have identified the basic microstructural unit whose percolation corresponds to the onset of DST in shear thickening suspensions, thanks to large scale 3D simulations of model suspensions.
Remarkably, these microstructural units can build a locally rigid network of frictional contacts and their cluster statistics follow the finite size scaling ansatz typical of critical phenomena and percolation theory. The percolation of these locally rigid structures is central in the stress transmission throughout the suspension and appears be at the origin of the giant fluctuations and the critical-like behaviors observed for DST. 
The snapshots in Figure \ref{fig:visualization} indeed show how the rigid structures identified through the $4$-neighbor particles change drastically with the global stress measured in the system. 

As our findings support the idea that a very specific self-organization of the suspension microstructure has to take place during DST, 
they call on experimental approaches that can recognize the locally rigid structures. For example, efforts in the past to study the conductivity of a suspension of particles, where the matrix fluid is conducting and the spherical particles are insulating, can distinguish between ordered and disordered states \cite{pascal}. However, this is not sufficient to distinguish between different $k$-neighbor particles, making it a challenge to identify the type of structure which is percolating. Experiments that can image stress fields \cite{Joriadze2013microscopic,Lin2016relating,Rathee2017localized} will likely pick out stress chains in compression. This is not necessarily enough to directly identify the $4$-neighbor particle network which appears to be crucial for DST, although the stability of a percolating stress chain may indicate it follows a connected path of $k=2$ neighbor particles which would be a subset of a network of $k=4$ neighbor particles, providing the necessary structure to support stress chains.
Developing capability to identify the mechanical constraints acting locally on particles in different part of the suspension contact network may be central to test the insight gained in this work. As extensive large scale simulations could allow, in the future, to more precisely determine the universality class from the percolation exponents, experimental rheological tests and scaling analysis of experimental flow curves \cite{Lin2016tunable,Ong2020,Ramaswamy2021} could complement the microscopic understanding developed here. Following the hierarchical self-organization of the $k$-neighbor particle structures under shear and identifying possible precursors of the rigidity percolation could also provide novel insight into shear thickening instabilities.



\section*{Simulation methods}
All simulations were performed with LAMMPS (Large-scale Atomic/Molecular Massively Parallel Simulator) \cite{Plimpton1995}. While generally known for molecular dynamics simulations, the LAMMPS code has specialized modules that allow for the modeling of soft sphere suspensions of Brownian particles. For this application, LAMMPS utilizes a simple physics-based discrete element method (DEM) model that simplifies the detailed flow behavior of the suspension solvent in exchange for computational efficiency. Hydrodynamic interactions between spheres are largely controlled by lubrication forces. Shear flow was imposed along the $x$ direction (with gradient along $z$) based on the Lees-Edwards boundary condition, {with an additional Stokesian drag force which causes particles to follow the imposed shear profile over time.} The robustness of this approach improves with increasing volume fraction of spheres as a less detailed knowledge of flow of the background fluid is needed. Indeed, it has been found that at volume fractions of approximately $40\,\%$ and higher, the flows produced are reasonably consistent with fully detailed simulations \cite{Feys2022}. This is due to the fact that, at higher volume fractions, the surfaces of the solid inclusions are close enough such that the lubrication forces dominate over the long-range hydrodynamics of the background fluid \cite{Ball1997}. A detailed description of this approach can be found in our previous paper \cite{Goyal2022}.

In addition to the hydrodynamic forces, frictional forces are included following the contact model of Mari et al \cite{Mari2014,Seto2013}. The steric repulsion between particles is modeled as a Hookean force, with a normal force of $F_N=kh$ that depends only on particle surface separation $h$ and spring constant $k$. This allows for particle contact, which is the criterion for activating frictional forces. To mimic a hard contact, we use a high spring constant of $k=k_0\dot{\gamma}$, with $k_0$ values between $10^7\,m{\tau_0}^{-2}$ and $10^8\,m{\tau_0}^{-2}$. With these parameters, we observe particle overlap $<3\,\%$, comparable to the criteria used in other recent simulations \cite{Morris2018}. The dependence on shear rate arises from the fact that a higher spring constant is needed to limit overlap at higher shear rates, due to collisions happening more frequently and at greater velocities. Alternatively, one could use a fixed, high value of $k$, corresponding to what is required for the highest shear rate, at all rates. However, this requires a smaller timestep to adequately resolve the collisions, and the variable spring constant approach was found to reduce computational cost while producing results that matched the results from the constant $k$ approach \cite{Morris2018}.

The frictional forces act tangentially to the particle contact and also follow a simple Hookean model. Any tangential displacement of particles after making contact, $\Delta r_t$, is acted upon by a restoring frictional force $F_t=k \Delta r_t$, { with the constraint that $F_t\leq F_N$ (Coulomb's friction law with a friction coefficient of $1$).} The contact/friction model we use is relatively simple compared to some recent studies \cite{Jamali2019,More2020}. However, all these models exhibit qualitatively similar behavior, and the simpler model also compares favorably with experiments \cite{Lee2020}. In this study, we focus mainly on the identification of a stress-bearing microstructural network with critical behavior, acknowledging that the exact values of the critical exponents obtained and their universality class is beyond the scope of this paper.

\section{Data analysis methods}
Simulations for all system sizes were run for at least 40 strain units to ensure sufficient sampling of the steady state flow. The data shown, except in cases where it is plotted as a function of $\gamma$, are averages over all sampled points with $\gamma>10$. The error bars correspond to the standard deviation of the average, and are smaller than the symbol size when not visible.

The stress is calculated from the particle contributions to the stress tensor, following \cite{Thompson2009}, as  

\begin{equation}
    \sigma_{\alpha \beta} = -\frac{1}{V}\sum_{i} \left[ \sum_{j \neq i} \left( \frac{1}{2} F_\alpha^{ij}r_\beta^{ij} \right) - mv_\alpha^i v_\beta^i \right]
    \label{eq:stressTotal}
\end{equation}
where $\alpha$ and $\beta$ can be $x$, $y$, or $z$ to generate the components of the stress tensor, $V$ is the system volume, $\vec{F}^{ij}$ and $\vec{r}^{ij}$ are the force and position vectors between particles $i$ and $j$, and $\vec{v}^i=\vec{v}^i_{\rm total}-\dot{\gamma}z \hat{x}$ is the deviation of the particle velocity from the flow profile set by the shear rate. The suspension viscosity is computed from the shear stress $\sigma_{xz}$ and shear rate $\dot{\gamma}$ as $\eta = \sigma_{xz}/\dot{\gamma}$.

For the visualizations in Fig. 5, the virial formulation of the stress tensor is broken up into particle contributions as:

\begin{equation}
    \sigma_{\alpha \beta}^i = -\sum_{j \neq i} \left[ \frac{1}{2} F_\alpha^{ij}r_\beta^{ij} \right] - mv_\alpha^i v_\beta^i
    \label{eq:stress}
\end{equation}

In the clustering analysis, particles are considered connected when they are in contact. In our simulation model, the contact can be exactly determined by a distance threshold between particle centers ($r_{ij}<d$ means particles overlap, experience steric repulsion, and are frictionally constrained). A cluster of particles is defined to be percolating in a direction if, when we split our system into thin ($<.45d$) slices along that axis, at least 1 particle from the cluster is in each slice. The percolation probability $R$ at any given $\dot{\gamma}$ is computed as the fraction of sampled $\gamma$ values at which a percolating cluster exists.

The probability P of a particle being in the percolating cluster is determined from the number of particles in the percolating cluster compared to the total $N$. $S$ and $S_{\rm max}$ are instead computed from all non-percolating clusters. $S_{\rm max}$ is the size of the largest of these clusters, averaged over $\gamma$. $S$ is the mean size of the clusters, given as:

\begin{equation}
    S = \sum_s \, s^2 n_s
\end{equation}
where $s$ is a cluster size and $n_s$ is the cluster size distribution (i.e. number of clusters of size $s$ normalized by $N$).
\section{Time scales}
\begin{figure}[h]
    \centering
    \includegraphics[width=.42\textwidth]{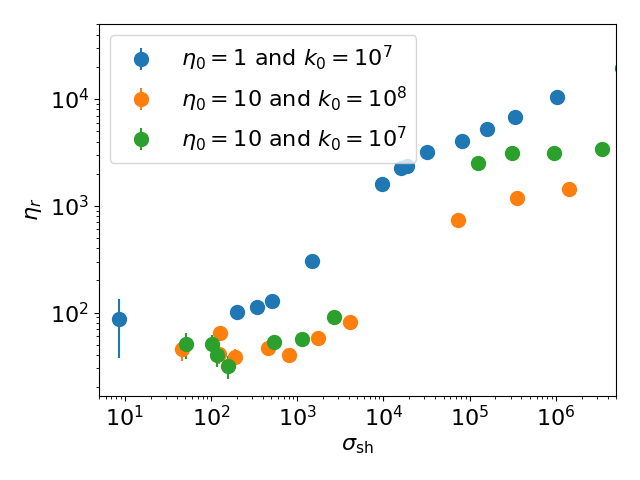}
    \caption{The relative viscosity plotted as a function of stress. The critical stress depends on parameters $\eta_0$ and $k_0$. }
    \label{fig:timescale1}
\end{figure}
\begin{figure}[htb]  
    \includegraphics[width=.42\textwidth]{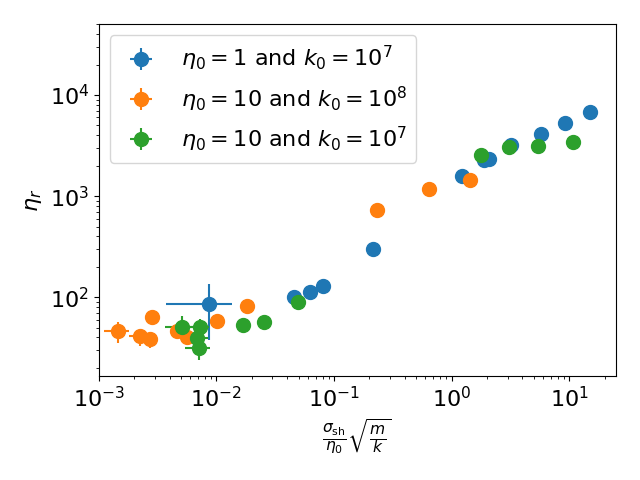}
    \caption{Rescaling the stress by $\eta_0$ and multiplying by the harmonic time scale $\sqrt{m/k}$ gives a dimensionless quantity which collapses the data to a single curve with a critical point close to 1.}
    \label{fig:timescale2}
\end{figure}
In order to test the simulation model, we performed simulations with a few different values of $\eta_0$ and $k_0$. Predictably, changing these parameters affected the critical rate at which DST was observed. We hypothesized that this might be explained by the changes in dimensionless numbers such as the Peclet number or Reynolds number (Re), but that does not seem to be the case in the flow regimes we studied. Although, it should be pointed out that increasing Re produces a gradual increase in viscosity for Re $>1$ . If we instead plot the relative viscosity vs the stress (Fig.~\ref{fig:timescale1}), we can see that there appears to be a shift in the critical stress that is coupled to the value of $\eta_0$. From this, we began to consider other time scales in the system.
  
One inverse time scale could be constructed from the ratio $\sigma/\eta_0$. This is the shear rate we expect from the Newtonian solvent without any particles. The natural rate related to $k_0$ is the resonant frequency of our Hookean model, $\sqrt{k/m}$. Plotting $\eta_r$ vs a dimensionless number given by the competition of these two rates ($\frac{\sigma}{\eta_0}\sqrt{\frac{m}{k}}$) produces a rather convincing data collapse, shown in Figs. ~\ref{fig:timescale2}. In addition, the DST transition appears quite close to $\frac{\sigma}{\eta_0}\sqrt{\frac{m}{k}}=1$, suggesting that shear-rates sufficiently high compared to the particle spring frequency are needed to allow the build-up of a percolating and stress-bearing frictional network that correspondingly increases the viscosity, effecting relevant time scales.

\end{document}